%% file: main.tex
\pdfoutput=1
\documentclass[11pt]{article}
\usepackage[letterpaper,margin=1in]{geometry}
\usepackage[T1]{fontenc}
\usepackage{lmodern}
\usepackage{microtype}
\usepackage{natbib}
\setcitestyle{authoryear,round,citesep={;},aysep={,},yysep={;}}


\usepackage{xcolor}
\usepackage[colorlinks=true,linkcolor=linkblue,citecolor=linkblue,urlcolor=linkblue,linktoc=page]{hyperref}
\hypersetup{pdftitle={EvolveMAS: A Unified Self-Evolution Framework for Decentralized Multi-Agent Systems},pdfauthor={Yalun Wu, Qiwei Ma, Bingbing Xu, Dan Zhang, Long Bai, Boyang Wang, Jiatong Li, Zheqi Lv, Shaozu Yuan, Peiying Wang, Jiawei Wang, An Zhang, Tat-Seng Chua}}
\usepackage{url}
\usepackage{xurl}

\usepackage{booktabs}
\usepackage{multirow}
\usepackage{makecell}
\usepackage{enumitem}
\usepackage{amsmath}
\usepackage{amssymb}
\usepackage{graphicx}
\usepackage{subcaption}
\usepackage{colortbl}
\usepackage{balance}
\usepackage{pifont}
\usepackage{listings}
\usepackage[most]{tcolorbox}
\usepackage{placeins}
\usepackage{tikz}
\usepackage{titlesec}
\usepackage{titletoc}
\titlecontents{section}[1.8em]{\addvspace{5pt}\sffamily\bfseries}{\contentslabel{1.8em}}{\hspace*{-1.8em}}{\hfill\contentspage}
\titlecontents{subsection}[4.2em]{\small}{\contentslabel{2.4em}}{\hspace*{-2.4em}}{\titlerule*[0.6pc]{.}\contentspage}
\usepackage{caption}

\definecolor{linkblue}{HTML}{1F5FD6}
\definecolor{cardgray}{HTML}{F2F3F5}
\definecolor{evolvegreen}{HTML}{17805E}
\titleformat{\section}{\sffamily\bfseries\Large}{\thesection}{0.8em}{}
\titleformat{\subsection}{\sffamily\bfseries\large}{\thesubsection}{0.8em}{}
\titleformat{\subsubsection}{\sffamily\bfseries\normalsize}{\thesubsubsection}{0.8em}{}
\titleformat{\paragraph}[runin]{\sffamily\bfseries\normalsize}{}{0pt}{}
\titlespacing*{\section}{0pt}{1.6ex plus .4ex minus .2ex}{0.9ex plus .2ex}
\titlespacing*{\subsection}{0pt}{1.3ex plus .3ex minus .2ex}{0.6ex plus .2ex}
\titlespacing*{\paragraph}{0pt}{1.0ex plus .3ex minus .2ex}{0.6em}
\newcommand{\linkrow}[2]{\noindent{\sffamily\bfseries\small #1:}\hspace{0.4em}{\small\url{#2}}\par\vspace{2pt}}

\newcommand{\Description}[1]{}

\lstdefinestyle{memdump}{
  basicstyle=\ttfamily\scriptsize,
  breaklines=true,
  breakatwhitespace=false,
  columns=fullflexible,
  keepspaces=true,
  showstringspaces=false,
  aboveskip=0pt,
  belowskip=0pt,
}
\newtcblisting{membox}[1]{
  enhanced,colback=gray!4,colframe=black!55,
  boxrule=0.4pt,arc=2pt,left=4pt,right=4pt,top=3pt,bottom=3pt,
  title={\footnotesize\bfseries #1},fonttitle=\small,
  coltitle=black,colbacktitle=gray!18,
  listing only,listing options={style=memdump},
}

\IfFileExists{latexml.sty}{\usepackage{latexml}}{\let\iflatexml\iffalse}
\newcommand{\paperTitle}{EvolveMAS: A Unified Self-Evolution Framework}
\newcommand{\paperSubtitle}{for Decentralized Multi-Agent Systems}
\newcommand{\paperAuthors}{%
  \paperAuthor{Yalun Wu}{1}\and%
  \paperAuthor{Qiwei Ma}{2}\and%
  \paperAuthor{Bingbing Xu}{3}\and%
  \paperAuthor{Dan Zhang}{1}\and%
  \paperAuthor{Long Bai}{3}\and%
  \paperAuthor{Boyang Wang}{4}\and%
  \paperAuthor{Jiatong Li}{1}\and%
  \paperAuthor{Zheqi Lv}{5}\and%
  \paperAuthor{Shaozu Yuan}{6}\and%
  \paperAuthor{Peiying Wang}{6}\and%
  \paperAuthor{Jiawei Wang}{1\dagger}\and%
  \paperAuthor{An Zhang}{7}\and%
  \paperAuthor{Tat-Seng Chua}{1}%
}
\newcommand{\paperAffiliations}{%
$^{1}$National University of Singapore\hspace{1.2em}$^{2}$Hunan University\\
$^{3}$Institute of Computing Technology, Chinese Academy of Sciences\\
$^{4}$Xingchen AGI, China Telecom Artificial Intelligence Technology Co., Ltd.\hspace{1.2em}$^{5}$Cornell University\\
$^{6}$Meituan\hspace{1.2em}$^{7}$University of Science and Technology of China\hspace{1.2em}$^{\dagger}$Corresponding author%
}
\newcommand{\paperAbstract}{%
Decentralized multi-agent systems (MAS) enable large language model agents to interact without centralized control. Agents need to adapt to other agents at test time and reuse accumulated experience to refine their interaction strategies for future interactions across cooperative, mixed-motive, and adversarial settings. However, existing approaches often develop adaptation mechanisms for particular interaction settings. This shared requirement raises a central research question: how can a decentralized MAS framework support both test-time adaptation and reliable experience reuse across various interaction settings? To this end, we first propose EvolveMAS, a unified self-evolution framework that supports experience-driven adaptation within and across episodes. EvolveMAS pairs each main agent with a dedicated MemoryAgent for test-time memory governance, which maintains isolated Structured Memory outside the main agent’s reasoning context. EvolveMAS organizes adaptation into two complementary processes: Within-Episode Online Adaptation and Cross-Episode Knowledge Evolution. During Within-Episode Online Adaptation, MemoryAgent updates episodic, semantic, and working memory from interaction evidence to inform subsequent actions. During Cross-Episode Knowledge Evolution, the main agent uses memory and episode outcomes to distill reusable interaction knowledge comprising partner profiles and interaction strategies for subsequent episodes.
Across five benchmarks and six evaluation configurations spanning the three interaction settings, EvolveMAS achieves the highest normalized average score of 77.2 among the compared methods. Ablations show that test-time memory governance and cross-episode knowledge reuse both improve performance. Interaction traces further illustrate how updated memory and distilled knowledge guide subsequent interactions. Overall, the experimental results support a unified self-evolution framework for experience-driven adaptation in a decentralized MAS environment without updating model parameters.%
}
\newcommand{\paperLinks}{%
\linkrow{Website}{https://wuyalun.github.io/EvolveMAS/}
\linkrow{Code}{https://github.com/WuYalun/EvolveMAS}
}
\iflatexml
  \newcommand{\paperAuthor}[2]{#1$^{#2}$}
  \title{\paperTitle{} \paperSubtitle}
  \expandafter\author\expandafter{\paperAuthors}
  \date{}
\else
  \newcommand{\paperAuthor}[2]{\mbox{#1$^{#2}$}}
\fi

\begin{document}
\thispagestyle{plain}
\iflatexml
\maketitle
\noindent\paperAffiliations\par
\begin{abstract}
\paperAbstract
\end{abstract}
\paperLinks
\else
\begin{tcolorbox}[enhanced,colback=cardgray,colframe=cardgray,boxrule=0pt,arc=9pt,
  left=16pt,right=16pt,top=16pt,bottom=14pt]
{\sffamily\bfseries\LARGE \paperTitle\\[2pt] \paperSubtitle\par}
\vspace{10pt}
{\sffamily\bfseries\small\raggedright
\renewcommand{\and}{, \hspace{0.45em}}\paperAuthors\par}
\vspace{8pt}
{\small\raggedright
\paperAffiliations\par}
\vspace{12pt}
{\small
\paperAbstract\par}
\vspace{10pt}
\begin{minipage}[b]{\linewidth}
\paperLinks
\end{minipage}
\end{tcolorbox}
\vspace{6pt}
\fi

\input{sections/Introduction}

\input{sections/Related}

\input{sections/Methods}

\input{sections/Experiments}

\input{sections/Conclusion}

\FloatBarrier

\enlargethispage{2\baselineskip}
\section*{Acknowledgements}
{\small\emergencystretch=1em
The framework and the per-benchmark experiment code are available at \url{https://anonymous.4open.science/r/artifact_7f3a/}, together with a README and a dependency list. Benchmarks and team composition are described in Section~\ref{sec:exp_setup}, the transfer setup of each source method family in Appendix~\ref{sec:transfer_setup}, and the full cross-setting comparison in Table~\ref{tab:matrix}.

This work evaluates language-model agents on public benchmarks. It involves no human subjects or personal data and releases no new dataset. The Werewolf evaluation involves simulated hidden-role play among model agents only, and we report village-side and wolf-side outcomes together so that deception ability is not presented on its own.\par}

\input{sections/Appendix}

\vspace{1em}
\renewcommand{\bibfont}{\small}
\setlength{\bibsep}{3pt plus 1pt}
\bibliographystyle{iclr2027_conference}
\bibliography{references}

\end{document}

%% file: sections/Introduction.tex
\section{Introduction}

\begin{figure}[tp]
    \centering
    \includegraphics[width=\textwidth]{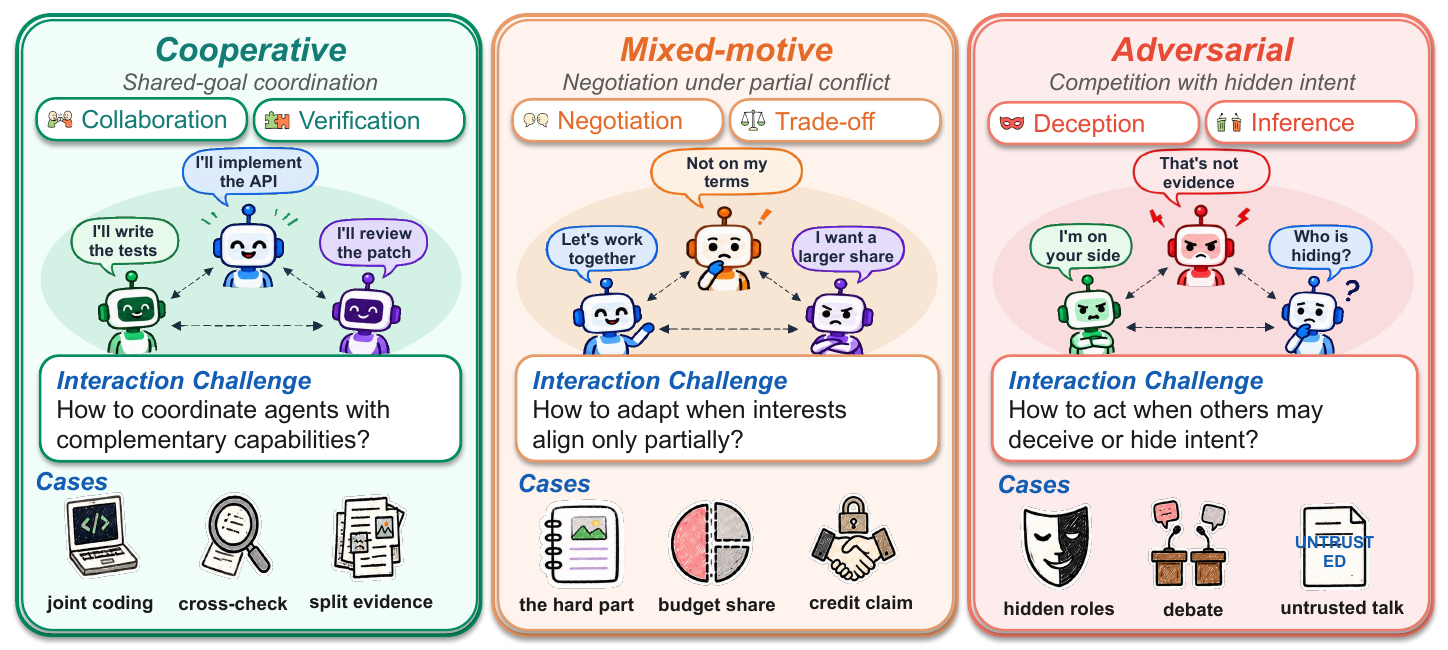}
    \caption{Three representative interaction regimes for decentralized MAS.}
    \label{fig:teaser}
    \Description{A three-panel teaser. The cooperative panel shows shared-goal coordination with joint coding, cross-checking, and split evidence. The mixed-motive panel shows negotiation under partial conflict, including budget share and credit claim. The adversarial panel shows competition
    with hidden intent, including hidden roles, misleading debate, and untrusted talk.}
\end{figure}

Large language model (LLM) agents are increasingly deployed to assist users with personal and professional tasks~\citep{staufer2026aiagentindex}.
Connecting agents deployed by different users creates opportunities to integrate their capabilities and support collaboration across users~\citep{chen2025ioa}.
Decentralized multi-agent systems (MAS) support such collaboration by allowing agents to act on local information and communicate with peers without a central controller~\citep{yang2025agentnet}.
From a game-theoretic perspective, multi-agent interactions are commonly categorized as cooperative, mixed-motive, or adversarial according to how agents' objectives align~\citep{busoniu2008comprehensive,zhang2021marloverview}. Cooperative settings involve shared goals, mixed-motive settings combine shared and conflicting interests, and adversarial settings involve opposing goals, as shown in Figure~\ref{fig:teaser}.

When interacting with unfamiliar partners, people usually learn through interaction, using that experience to adjust immediate responses and guide future interactions. Agents similarly need to learn online from interaction experience to build profiles of other agents and refine their interaction strategies for subsequent interactions~\citep{he2016opponent,albrecht2017hypothetical}. This raises a central research question: \emph{how can a unified mechanism support test-time adaptation across cooperative, mixed-motive, and adversarial settings?}
Experience-driven self-evolution methods provides an insight to such adaptation by using feedback and accumulated experience to improve behavior without heavy parameter updates~\citep{wang2024awm,tang2025chemagent,ouyang2025reasoningbank}.
However, recent research shows that accumulating long interaction histories in the reasoning context creates context bloat that can affect task performance~\citep{du2025context}.
Retaining inaccurate accumulated experience introduces memory pollution that can misguide subsequent interactions~\citep{xiong2026memorymanagement}. These two limitations further raise the question: \emph{how can self-evolution decentralized agents achieve reliable experience reuse across three interaction settings?}

To address these questions, we propose \textbf{EvolveMAS}, a unified self-evolution framework that enables decentralized agents to learn from interaction experience across cooperative, mixed-motive, and adversarial settings. EvolveMAS pairs each main agent with a dedicated MemoryAgent for test-time memory governance to address context bloat and memory pollution. The MemoryAgent maintains Structured Memory outside the main agent’s reasoning context, organizing interaction evidence into Episodic, Semantic, and Working Memory. During Within-Episode Online Adaptation phase, the MemoryAgent updates Structured Memory as interactions unfold to inform subsequent actions. During Cross-Episode Knowledge Evolution phase, the main agent uses interaction outcomes to refine reusable Knowledge comprising Partner Profiles and Interaction Strategies for future better interactions.

\begin{figure}[!t]
  \centering
  \includegraphics[width=0.8\textwidth]{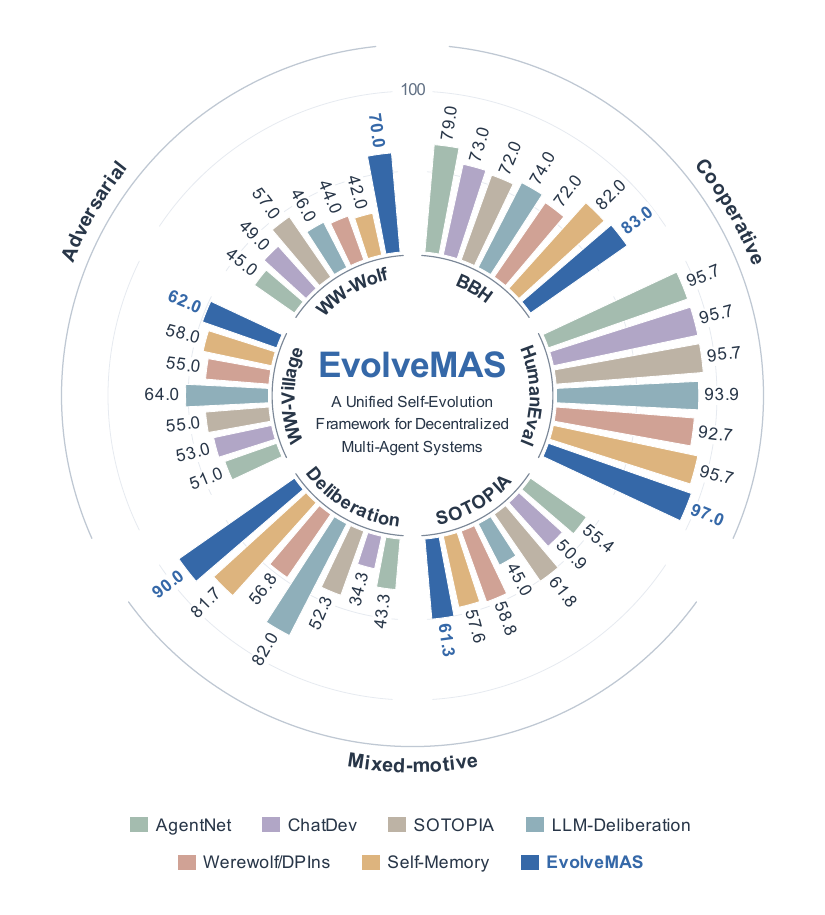}
  \caption{Results at a glance: the seven methods on the six configurations of
  Table~\ref{tab:matrix}. Compare bar lengths only within a sector; EvolveMAS is dark blue.}
  \label{fig:radar}
\end{figure}

We evaluate EvolveMAS on five benchmarks spanning cooperative, mixed-motive, and adversarial settings. Under a shared self-evolution protocol, EvolveMAS achieves the highest normalized average score of 77.2 (Figure~\ref{fig:radar}), exceeding the strongest setting-specific baseline by 9.7 points. Ablation studies demonstrate that test-time memory governance improves interaction performance, with additional improvements from cross-episode Knowledge reuse. Additional analyses show that online memory updates improve question-answering accuracy over offline updates and that test-time memory governance reduces active-context growth. 

Our contributions can be summarized as follows:
\begin{itemize}[topsep=2pt,itemsep=1pt,parsep=0pt]
    \item We propose EvolveMAS, a unified self-evolution framework that supports experience-driven adaptation in decentralized MAS across cooperative, mixed-motive, and adversarial settings through Within-Episode Online Adaptation and Cross-Episode Knowledge Evolution.

    \item We introduce a dedicated MemoryAgent for test-time memory
    governance that maintains Structured Memory to support the main agent's reasoning during interaction and cross-episode Knowledge reuse through Partner Profiles and Interaction Strategies.

    \item We evaluate EvolveMAS on five benchmarks across the three interaction settings and demonstrate the effectiveness of test-time memory governance and cross-episode Knowledge reuse through ablation studies. 
\end{itemize}

%% file: sections/Related.tex
\section{Related Work}

\paragraph{Decentralized Multi-Agent Systems.}

Research on decentralized multi-agent systems has explored coordination and
adaptation across diverse interaction regimes. Game-theoretic studies
commonly classify multi-agent interactions as cooperative, mixed-motive,
or adversarial~\citep{yang2020gameperspective,wong2022deepmarl}, while recent
LLM-based MAS instantiate these regimes through cooperative coding and
reasoning~\citep{chatdev2024,du2024multiagentdebate,liang2023madebate},
mixed-motive social interaction~\citep{zhou2024sotopia,sotopia_pi2024}, and
adversarial strategic reasoning~\citep{werewolf_dpins,agarwal2025wolf,jiang2026omar}.
To support adaptation during interaction, classical opponent modeling infers
other agents' behavioral patterns from interaction history to guide
decision-making~\citep{carmel1996opponent,he2016opponent}.
More broadly, recent MAS adaptation methods adjust system topology,
workflows, roles, communication structures, or agent
policies~\citep{yang2025agentnet,hu2024evomac,hu2025adas,li2025aflow,
zhuge2024gptswarm,qian2024macnet,chen2023agentverse,wang2024mixture,
chen2024optima,liao2025marft,liu2025magrpo,xue2025comas,maporl2025}.
These approaches improve performance on specific coordination or adaptation
objectives and are typically tailored to particular interaction settings,
system structures, or training procedures.
In contrast, EvolveMAS maintains the same decentralized system organization
and evolution framework across these interaction regimes.

\paragraph{Self-Evolution.}

Recent work explores test-time self-evolution by reusing interaction
experience and feedback without updating model parameters.
Reflexion~\citep{shinn2023reflexion}, ExpeL~\citep{zhao2023expel}, and
Self-Refine~\citep{madaan2023selfrefine} demonstrate experience-driven
improvements on single-agent tasks, while EvoMAC~\citep{hu2024evomac} and
AgentNet~\citep{yang2025agentnet} extend iterative adaptation to multi-agent
systems.
Related studies also investigate self-generated data and self-play as
mechanisms for agent evolution~\citep{zelikman2022star,wang2022selfinstruct,
chen2024spin,huang2025rzero,zhao2025absolutezero,chen2025mae}.
However, existing self-evolution methods are primarily designed for specific
tasks or interaction settings.
To support self-evolution across interaction settings, EvolveMAS couples
within-episode adaptation with cross-episode Knowledge evolution through a
shared test-time evolution mechanism for decentralized MAS.

\paragraph{Agent Memory.}

Agent memory enables experience to persist beyond a single interaction
context~\citep{park2023generative,sumers2024cogarch}.
MemGPT~\citep{packer2023memgpt} and
MemoryBank~\citep{zhong2023memorybank} introduce persistent memory, while
Voyager~\citep{wang2023voyager}, Agent Workflow Memory~\citep{wang2024awm},
and SkillWeaver~\citep{zheng2025skillweaver} accumulate reusable skills and
workflows.
More recent work explores structured memory organization and test-time
memory reasoning~\citep{xu2025amem,chhikara2025mem0,
ouyang2025reasoningbank,zhou2025memento,yan2025memoryr1}.
However, in long-horizon decentralized interactions, retaining accumulated
memory in the main agent's context can lead to context expansion,
attention dilution, and memory
pollution~\citep{liang2026mcma,zhang2026memoryaction,wu2026remember,
wan2026compass}.
To address these challenges, EvolveMAS pairs each main agent with a dedicated
external MemoryAgent.
This MemoryAgent maintains an isolated Structured Memory and provides a
governed interaction state to support both adaptation timescales.

%% file: sections/Methods.tex
\section{EvolveMAS: Memory-Governed Self-Evolution}
\label{sec:method}

We propose EvolveMAS, a decentralized test-time self-evolution framework for
heterogeneous multi-agent interactions.
The framework combines per-agent external memory governance with a shared
Execute--Evaluate--Evolve (EEE) process.
Together, these components support within-episode adaptation and
cross-episode Knowledge evolution without updating model parameters.

\subsection{Framework Overview}
\label{sec:method_overview}

Consider a decentralized multi-agent system (DMAS) with main agents
$\mathcal{A}=\{a_i\}_{i=1}^{n}$ operating over a continuous stream of
heterogeneous interactions organized into episodes.
In episode $e$, the agents receive a task description $x_e$ and interact
autonomously over $T_e$ turns.
Each agent selects its actions based on local observations, while the task
determines which agents interact and when.
The resulting interaction record forms the trajectory
$\tau_e=(\tau_{e,1},\dots,\tau_{e,T_e})$, where $\tau_{e,t}$ contains the
messages and environment feedback produced at turn $t$.
Each agent observes its own local view of $\tau_e$, and the episode
concludes with an interaction outcome $o_e$.

\begin{figure*}[tp]
    \centering
    \includegraphics[width=\textwidth]{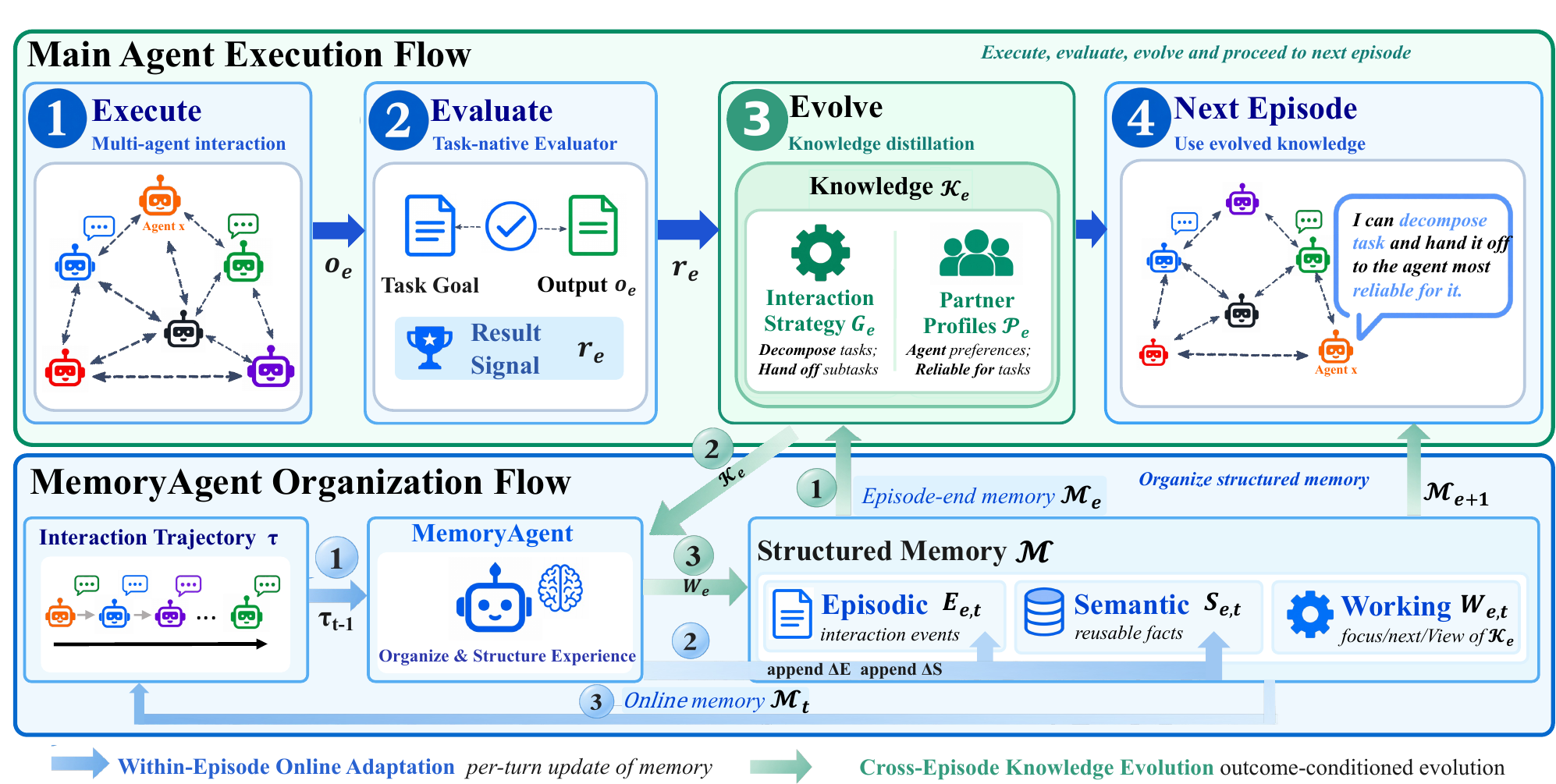}
    \caption{
    Overview of EvolveMAS. The upper lane is the main agent's
    Execute--Evaluate--Evolve cycle and the lower lane the MemoryAgent that
    maintains Structured Memory. Blue arrows are within-episode updates,
    green arrows cross-episode evolution.
    }
    \label{fig:method}
\end{figure*}

EvolveMAS preserves this decentralized organization
(Figure~\ref{fig:method}).
Each main agent $a_i$ is paired one-to-one with a dedicated MemoryAgent
$m_i$ and an isolated memory, allowing accumulated experience to inform
the main agent's subsequent behavior.
The main agent handles task reasoning, communication, and environment
interaction.
The corresponding MemoryAgent maintains Structured Memory and Interaction
Knowledge outside the main agent's reasoning context, drawing on interaction
evidence observed by that main agent.
All MemoryAgents share the same memory schema and governance protocol,
while their contents evolve independently based on their respective
main agents' local interaction histories.

Based on the interaction record, EvolveMAS represents experience at three
levels of abstraction.
The \emph{Trajectory} $\tau_e$ consists of raw interaction evidence
generated during task execution.
Building on this evidence, EvolveMAS maintains \emph{Structured Memory}
$M_{e,t}$ as a compact state updated during interaction.
At a higher level, it maintains \emph{Interaction Knowledge} $K_e$ to
capture reusable regularities across episodes.
For clarity, the subsequent formulation adopts the perspective of a focal
main agent $a=a_i$, with the agent index omitted from its memory and
Interaction Knowledge states.

Let $M_e\equiv M_{e,T_e}$ denote the episode-end Structured Memory.
Here, $o_e$ denotes the observed interaction outcome, and $r_e$ denotes
the corresponding task-native result signal.
The EEE process is then formulated at the episode level as
\begin{equation}
\begin{aligned}
\textsc{Execute}: \quad
    &(\tau_e,M_e,o_e)
    =
    \operatorname{Exec}
    \bigl(x_e,\mathcal{A},M_{e,0},K_{e-1}\bigr),
\\
\textsc{Evaluate}: \quad
    &r_e
    =
    \operatorname{Eval}
    \bigl(x_e,\tau_e,o_e\bigr),
\\
\textsc{Evolve}: \quad
    &K_e
    =
    D_{a}
    \bigl(M_e,r_e;K_{e-1}\bigr).
\end{aligned}
\label{eq:eee}
\end{equation}
Here, $\operatorname{Eval}$ denotes the task-native outcome evaluator.
Accordingly, $r_e$ may vary in form across interaction settings while
providing a common interface for outcome-conditioned evolution.
The operator $D_a$ distills Interaction Knowledge from governed
Structured Memory, the observed result, and previously accumulated
Interaction Knowledge.
The resulting $K_e$ comprises reusable Interaction Strategies
and Partner Profiles.
Both \textsc{Execute} and \textsc{Evolve} use the previous episode's
Knowledge $K_{e-1}$.
Specifically, \textsc{Execute} uses it to guide interaction, whereas
\textsc{Evolve} incorporates new experience to refine it.
In contrast, \textsc{Evaluate} computes $r_e$ from $x_e$, $\tau_e$,
and $o_e$, without directly accessing accumulated memory or Knowledge.
The following subsections detail the memory and Knowledge
update mechanisms.

\subsection{Within-Episode Online Adaptation}
\label{sec:online_memory}

During \textsc{Execute} in Eq.~\ref{eq:eee}, main agents interact with
peers and the environment.
As interaction proceeds, their MemoryAgents continually update Structured
Memory based on newly generated interaction evidence.
At turn $t$, action generation and memory updating are formulated as
\begin{equation}
\begin{aligned}
(y_{e,t},\Delta\tau_{e,t})
&\sim
\pi_{a}
\left(
\cdot
\mid
[x_e;M_{e,t-1};K_{e-1}]
\right),
\\
M_{e,t}
&=
\operatorname{MA}_{\mathrm{online}}
\left(
M_{e,t-1},
\Delta\tau_{e,t}
\right),
\end{aligned}
\label{eq:online_update}
\end{equation}
where $y_{e,t}$ denotes the generated response or environment action,
and $\Delta\tau_{e,t}$ denotes the newly observed interaction evidence
at turn $t$.

Structured Memory comprises three complementary regions,
\begin{equation}
M_{e,t}
=
\left(
E_{e,t},
S_{e,t},
W_{e,t}
\right),
\label{eq:structured_memory}
\end{equation}
where $E_{e,t}$, $S_{e,t}$, and $W_{e,t}$ denote Episodic, Semantic, and
Working Memory~\citep{tulving1972episodic,baddeley1974working}.

\paragraph{Episodic Memory.}
$E_{e,t}$ records concrete interaction experiences, including the events
that occurred, the agents or sources involved, and the outcomes that
followed.
Together, these records provide an interaction-grounded history from
which regularities can subsequently be inferred.

\paragraph{Semantic Memory.}
$S_{e,t}$ maintains reusable facts and interaction regularities inferred
from experience.
Depending on the interaction, these may include task-relevant facts,
recurring coordination patterns, observations about interaction partners,
or other stable information that supports subsequent reasoning.

\paragraph{Working Memory.}
$W_{e,t}$ maintains a compact state relevant to the ongoing interaction.
This state includes the current focus, next-step information, and
task-relevant views of accumulated memory and Interaction Knowledge.
As the interaction context changes, Working Memory is refreshed
accordingly.

At the representation level, new events are appended to Episodic Memory,
reusable regularities are incorporated into Semantic Memory, and Working
Memory is refreshed.
These conceptual update rules do not prescribe a specific storage
implementation.
\looseness=-1

The updates in Eq.~\ref{eq:online_update} form a recurrent feedback loop
within \textsc{Execute}.
Evidence observed at turn $t$ updates Structured Memory, which conditions
the acting policy $\pi_a$ from turn $t+1$ onward.
This loop enables Within-Episode Online Adaptation.
Sec.~\ref{sec:cross_episode} formalizes both adaptation timescales
in Eq.~\ref{eq:two_timescales}.

The term \emph{within-episode} specifies the timescale of this feedback loop.
Episodic and Semantic information may persist and accumulate across
episodes, whereas Working Memory is refreshed for the ongoing interaction.

\subsection{Cross-Episode Knowledge Evolution}
\label{sec:cross_episode}

After episode $e$ concludes, \textsc{Evaluate} produces a task-native
result signal $r_e$ from the observed interaction outcome.
This signal serves as a protocol-level interface for conditioning
subsequent evolution without requiring a shared metric space across
heterogeneous interactions.
Conditioned on $r_e$, the main agent distills reusable interaction
regularities from the governed Structured Memory $M_e$ and previously
accumulated Interaction Knowledge $K_{e-1}$.
The evolved Interaction Knowledge comprises two complementary components,
\begin{equation}
K_e=(G_e,P_e),
\label{eq:knowledge}
\end{equation}
which are described below.

\paragraph{Interaction Strategies.}
$G_e$ captures reusable regularities that guide behavioral adaptation
across recurring interaction situations, including coordination,
verification, negotiation, and trust calibration.

\paragraph{Partner Profiles.}
$P_e=\{p_{e,j}\}_{j\neq i}$ captures partner-specific regularities, including
recurring behavioral tendencies, inferred capabilities, and reliability, and
is updated as new interaction evidence accumulates.
\looseness=-1

In subsequent episodes, the evolved Interaction Knowledge conditions \textsc{Execute}
together with the current Structured Memory. Combined with the online memory
updates in Sec.~\ref{sec:online_memory}, EvolveMAS therefore operates across
two coupled adaptation timescales:
\begin{equation}
\underbrace{
\Delta\tau_{e,t}
\rightarrow
M_{e,t}
\rightarrow
\textsc{Execute}_{t+1}
}_{\text{Within-Episode Online Adaptation}}
\qquad
\underbrace{
(M_e,r_e,K_{e-1})
\rightarrow
K_e
\rightarrow
\textsc{Execute}_{\mathrm{future}}
}_{\text{Cross-Episode Knowledge Evolution}}.
\label{eq:two_timescales}
\end{equation}
Together, within-episode evidence integration and cross-episode knowledge
reuse form a unified self-evolution framework for heterogeneous
interactions without updating model parameters.

%% file: sections/Experiments.tex
\section{Experiments}

\subsection{Experiment Setting}
\label{sec:exp_setup}

We evaluate EvolveMAS on five representative benchmarks spanning three
interaction settings.
The cooperative setting includes BBH~\citep{suzgun2023bbh} and
HumanEval~\citep{chen2021humaneval}, covering multi-agent reasoning and
collaborative coding under shared goals.
The mixed-motive setting includes
LLM-Deliberation~\citep{abdelnabi2024negotiation} and
SOTOPIA~\citep{zhou2024sotopia}, covering structured negotiation and
open-ended social interaction with partially aligned objectives.
The adversarial setting uses One Night Ultimate
Werewolf~\citep{werewolf_dpins}, a hidden-role game involving deception,
trust calibration, and strategic reasoning.

The main experiments use heterogeneous teams spanning model families
and capability tiers to support Partner Profile modeling.
Team size and composition follow each benchmark's native protocol
and difficulty.
Each main agent is paired with an isolated GPT-4o-mini MemoryAgent.
We evaluate five source method families under six configurations,
including benchmark-native and cross-setting comparisons.
Appendices~\ref{sec:transfer_setup} and~\ref{sec:transfer_robustness}
detail team composition, transfer construction, and additional analyses.
\emph{Self-Memory} retains the same interaction structure but manages
memory within the main agent's task-reasoning context.

\subsection{Main Results across Interaction Settings}

Table~\ref{tab:matrix} compares EvolveMAS with setting-specific
baselines transferred across interaction settings and with Self-Memory.
Under a shared self-evolution protocol, EvolveMAS ranks first in
four evaluation configurations and second in the remaining two.
Its normalized average score of 77.2 exceeds the strongest
setting-specific baseline by 9.7 points and Self-Memory by 7.7 points.

\begin{table*}[tp]
\caption{Cross-setting results. WW-V and WW-W report village win
and wolf deception success in one-sided Werewolf audits.
Shading identifies source settings. Bold and underlined values denote
the best and second-best results in each column, respectively, including ties.
Avg.\ is the mean of six percentage-scaled metrics:
SOTOPIA is multiplied by 100 and Deliberation by $100/6$.}
\label{tab:matrix}
\centering
\footnotesize
\setlength{\tabcolsep}{4pt}
\renewcommand{\arraystretch}{1.15}
\begin{tabular*}{\textwidth}{@{\extracolsep{\fill}}l|cc|cc|cc|c@{}}
\toprule
& \multicolumn{2}{c|}{Cooperative} & \multicolumn{2}{c|}{Mixed-motive} & \multicolumn{2}{c|}{Adversarial} & \multirow{2}{*}{Avg.} \\
\cmidrule(lr){2-3} \cmidrule(lr){4-5} \cmidrule(lr){6-7}
Method & BBH & HumanEval & SOTOPIA & Deliberation & WW-V & WW-W & \\
\midrule
AgentNet        & \cellcolor{gray!15}79 & \underline{95.7} & 0.554 & 2.60 & 51.0 & 45.0 & 61.6 \\
ChatDev         & 73 & \cellcolor{gray!15}\underline{95.7} & 0.509 & 2.06 & 53.0 & 49.0 & 59.3 \\
SOTOPIA         & 72 & \underline{95.7} & \cellcolor{gray!15}\textbf{0.618} & 3.14 & 55.0 & \underline{57.0} & 65.6 \\
LLM-Deliberation & 74 & 93.9 & 0.450 & \cellcolor{gray!15}\underline{4.92} & \textbf{64.0} & 46.0 & 67.5 \\
Werewolf/DPIns  & 72 & 92.7 & 0.588 & 3.41 & \cellcolor{gray!15}55.0 & \cellcolor{gray!15}44.0 & 63.2 \\
\midrule
Self-Memory     & \underline{82} & \underline{95.7} & 0.576 & 4.90 & 58.0 & 42.0 & \underline{69.5} \\
EvolveMAS       & \textbf{83} & \textbf{97.0} & \underline{0.613} & \textbf{5.40} & \underline{62.0} & \textbf{70.0} & \textbf{77.2} \\
\bottomrule
\end{tabular*}
\end{table*}

\paragraph{Cooperative setting.}
EvolveMAS achieves 83.0 on BBH, exceeding the native AgentNet
baseline of 79.0 by four percentage points.
On HumanEval, it achieves 97.0, compared with 95.7 for both
ChatDev and Self-Memory.
Appendix Table~\ref{tab:bbh_long} further examines performance
under an alternative BBH model configuration.

\paragraph{Mixed-motive setting.}
On LLM-Deliberation, EvolveMAS achieves the highest average agreement
of 5.40 parties, compared with 4.92 for the native baseline.
On SOTOPIA, it ranks second at 0.613, close to the native baseline
of 0.618 and 0.037 above Self-Memory.
Both settings provide partner-specific and negotiation regularities
that can be reused through cross-episode evolution.

\paragraph{Adversarial setting.}
Because adversarial roles have conflicting objectives, the one-sided
audits apply EvolveMAS to one side at a time.
On WW-V, EvolveMAS scores 62.0, compared with the highest score of
64.0 achieved by LLM-Deliberation.
On WW-W, it achieves the highest score of 70.0, exceeding the best
competing method by 13 points.
Appendix~\ref{sec:directional_audit} reports both one-sided conditions
and the two symmetric references on a common village-win scale.
\looseness=-1

Detailed metrics and per-benchmark breakdowns for
Table~\ref{tab:matrix} are provided in
Appendices~\ref{sec:exp_details} and~\ref{sec:additional_within_setting}.

\begin{figure*}[tp]
  \centering
  \includegraphics[width=\textwidth]{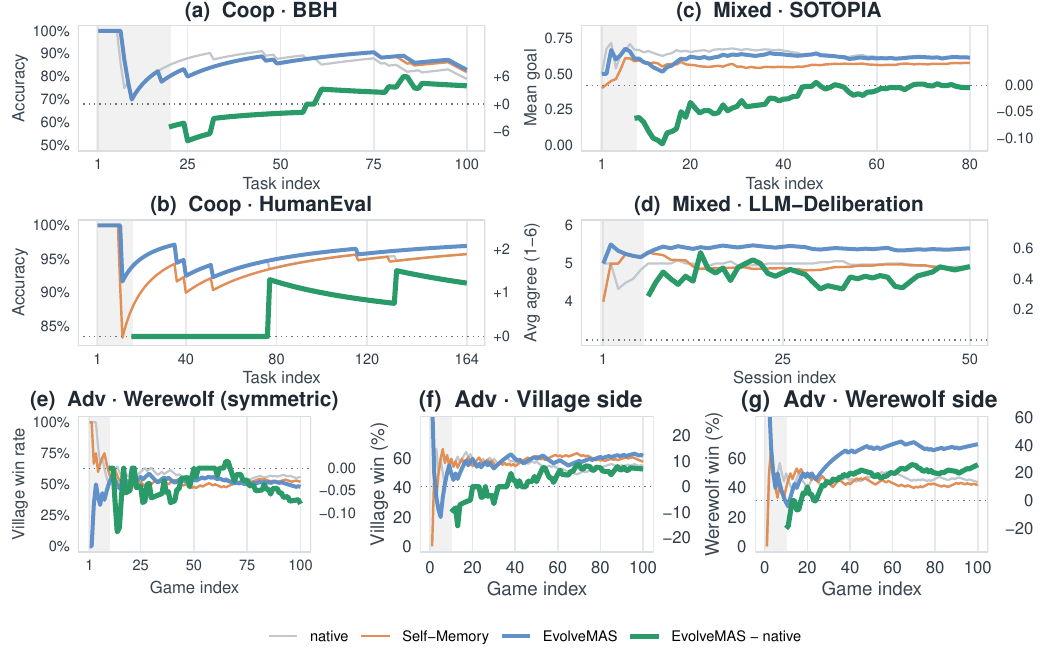}
  \caption{Within-session cumulative evolution. The cooperative (a--b) and
  mixed-motive (c--d) blocks are stacked side by side, and the adversarial row
  reports symmetric Werewolf (e) with the one-sided ablations, scored as village
  win (f) and as wolf deception success (g). Lines are cumulative means; the thicker green line is
  the paired difference (EvolveMAS $-$ native), on the right-hand axis, with
  its zero dotted and its scale set independently in each panel. Shading marks
  each panel's early, unstable tasks or games.}
  \label{fig:evolution}
\end{figure*}

\subsection{Cross-Setting Evolution}
\label{sec:evolution}

We examine how performance evolves as interaction experience accumulates.
Figure~\ref{fig:evolution} presents cumulative performance over tasks,
sessions, or games, grouped by interaction setting.
The trajectories compare EvolveMAS with the native baseline and
Self-Memory to assess whether accumulated experience continues to
influence subsequent behavior under the same evolution protocol.

In the cooperative setting, the relative performance of EvolveMAS on BBH
improves from an initial deficit to a sustained advantage over the later
tasks, ending at $+4$ percentage points.
On HumanEval, performance remains comparable to the native baseline
for roughly the first half of the task stream, followed by a modest
advantage that ends at approximately $+1.2$ percentage points.
In the mixed-motive setting, SOTOPIA shows an overall upward trend in
relative performance, with the early deficit narrowing toward parity
despite intermediate fluctuations.
On LLM-Deliberation, the cumulative advantage is positive throughout
the displayed interval and persists through the final session.
This pattern is consistent with adaptation taking effect early and
providing sustained benefits in subsequent interactions.
In the adversarial setting, symmetric Werewolf tracks village-side
performance as both opposing sides evolve.
The one-sided evaluations show positive advantages that persist over
the later games, with larger gains on the wolf side.

The trajectories in Figure~\ref{fig:evolution} indicate that accumulated
interaction experience supports continual evolution across cooperative,
mixed-motive, and adversarial settings under a shared EvolveMAS protocol.
Although trajectory shapes vary with the interaction structure, the
self-evolution process, memory representation, and update schedule
remain unchanged.

\subsection{Ablating the Two Evolution Mechanisms}
\label{sec:ablation}

To distinguish the contributions of the two mechanisms, the evaluation
requires both emerging information within sessions and recurring partners
across sessions.
These conditions support within-episode adaptation and cross-episode
Knowledge reuse, respectively.
LLM-Deliberation meets both conditions: participants reveal their positions
over successive rounds, and the same six parties participate in every session.
Its graded outcomes also allow the contribution of each mechanism to be
measured.
We report average agreement, measured as the mean number of parties accepting
the final proposal, and 5/6 consensus on that proposal.
The unanimity rate measures the fraction of sessions in which at least one
lead-party proposal is accepted by all six parties.

\begin{figure}[tp]
  \centering
  \includegraphics[width=\linewidth]{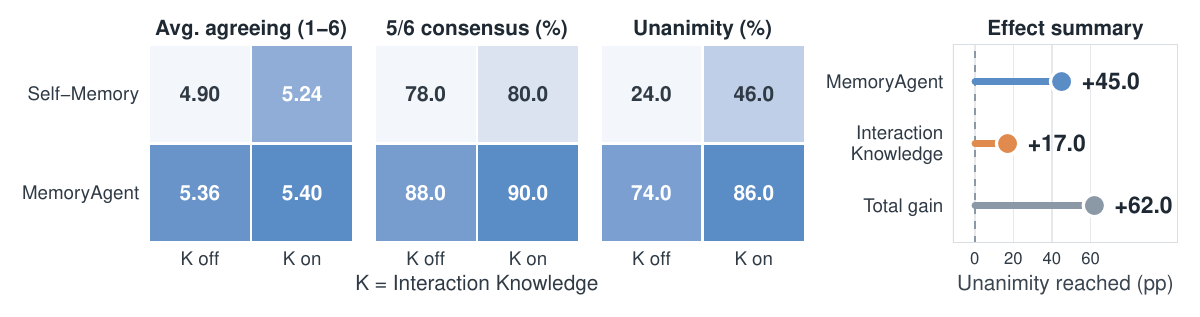}
  \caption{LLM-Deliberation ablation.
Rows vary memory governance; columns vary the availability of
cross-episode Knowledge $K_e=(G_e,P_e)$.
Each cell reports one run of 50 sessions; color scales are
panel-specific.
The effect summary reports average marginal effects on unanimity
and the total gain.}
  
  \label{fig:ablation}
\end{figure}

We independently vary memory governance
(Self-Memory vs.\ external MemoryAgent) and the availability of
cross-episode Interaction Knowledge
($K=\varnothing$ vs.\ $K_e=(G_e,P_e)$).
This yields the $2\times2$ factorial design in Figure~\ref{fig:ablation}.
The lower-right condition in each metric panel combines external memory
governance with cross-episode Interaction Knowledge and corresponds to
full EvolveMAS.
\looseness=-1

Unanimity rates differ across all four conditions.
Replacing Self-Memory with an external MemoryAgent increases unanimity
from 24.0\% to 74.0\% without cross-episode Knowledge and from
46.0\% to 86.0\% with it.
These gains are consistent with updated interaction states helping agents
respond to positions revealed during deliberation.
Adding $K_e$ further increases unanimity from 74.0\% to 86.0\% with an
external MemoryAgent and from 24.0\% to 46.0\% with Self-Memory.
These improvements support the value of reusing negotiation regularities
across episodes.
Average agreement follows the same ordering (4.90, 5.24, 5.36, 5.40).
Appendix~\ref{sec:ablation_scope} reports additional metrics, which follow
this ordering except for final-proposal unanimity.
Appendix~\ref{sec:ablation_all} extends the ablation to three additional
benchmarks.
\looseness=-1

\subsection{Case Study}
\label{sec:case}

\begin{figure*}[tp]
  \centering
  \includegraphics[width=\textwidth]{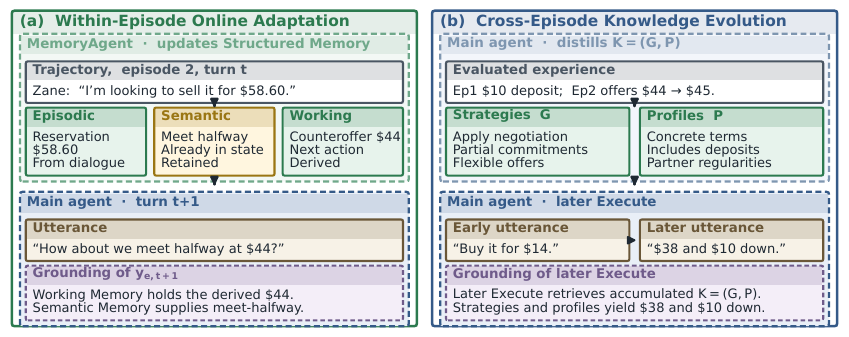}
  \caption{Instrumented SOTOPIA example illustrating
  Within-Episode Online Adaptation (a) and Cross-Episode Knowledge
  Evolution (b). In (a), entries written at the current turn are green and
  carried-over entries are yellow.}
  \Description{A two-panel case study showing within-episode Structured
  Memory updates from newly observed negotiation evidence and cross-episode
  distillation of Interaction Strategies and Partner Profiles for subsequent
  reuse.}
  \label{fig:sotopia_case}
\end{figure*}

Figure~\ref{fig:sotopia_case} illustrates the two adaptation processes
in Eq.~\ref{eq:two_timescales} through an instrumented EvolveMAS
trajectory in SOTOPIA negotiation.

\paragraph{Within-Episode Online Adaptation.}
In episode~2, the seller states an asking price of \$58.60
(Figure~\ref{fig:sotopia_case}a).
The MemoryAgent incorporates this evidence into Structured Memory
while retaining the existing meet-halfway regularity.
The resulting Working Memory contains a \$44 counteroffer, which the
buyer uses on the next turn.
This sequence illustrates how newly observed evidence informs subsequent
decisions within the same episode.
\looseness=-1

\paragraph{Cross-Episode Knowledge Evolution.}
Outcome-conditioned distillation produces reusable Interaction
Strategies and Partner Profiles across episodes
(Figure~\ref{fig:sotopia_case}b).
Experiences involving a \$10 deposit and a counteroffer revised
from \$44 to \$45 yield negotiation regularities.
The Partner Profile captures the counterpart's responsiveness
to concrete terms.
The resulting Interaction Knowledge guides subsequent offer
formulation during \textsc{Execute}.

Appendix~\ref{sec:qualitative} provides further examples across
all three interaction settings.

\subsection{Additional Analyses}

System-level analyses show stable performance across the evaluated
team sizes and favorable performance relative to static aggregation.
Further diagnostics show stable short-horizon performance and
lower per-episode cost and acting-context occupancy than retaining
raw interaction histories as experience accumulates
(Appendix~\ref{app:system_diagnostics}).
Analyses of Knowledge accumulation indicate that stronger acting
models distill more transferable Interaction Knowledge
(Appendix~\ref{sec:model_heterogeneity}).
Finally, a long-horizon memory ablation shows higher overall
question-answering accuracy with online MemoryAgent updates than
with offline updates or self-managed memory
(Appendix~\ref{app:locomo}).

%% file: sections/Conclusion.tex
\section{Conclusion}
We presented EvolveMAS, a unified self-evolution framework for decentralized multi-agent systems (DMAS). Evaluations on five benchmarks demonstrate its applicability to cooperative, mixed-motive, and adversarial settings, while ablations highlight the value of test-time memory governance and cross-episode Knowledge reuse. Together, these results support experience-driven adaptation across the three interaction settings under a shared self-evolution framework without model parameter updates.
Designed as a plug-in extension to existing agents, EvolveMAS supports decentralized interaction while preserving each agent’s autonomy. This provides a basis for integrating the capabilities of agents deployed by different users without centralized control. Building on this paradigm, future work can investigate empirical scaling laws by varying agent capability, population size, and accumulated experience. Such studies would clarify the design trade-offs among DMAS reasoning performance, computational cost, and safety.

%% file: sections/Appendix.tex
\appendix
\FloatBarrier
\vspace{1.2em}
\startcontents[appendix]
\noindent{\sffamily\bfseries\Large Appendix\par}
\vspace{8pt}
\printcontents[appendix]{}{1}{\setcounter{tocdepth}{2}}
\clearpage

\section{Experimental Protocols}
\label{sec:exp_details}
\suppressfloats[t]

The main text reports the cross-setting results and their interpretation.
This section instead documents how the transferred method families are
instantiated and how the full Deliberation metric suite is computed.

\subsection{Transfer Construction}
\label{sec:transfer_setup}

To make the transfer rows in Table~\ref{tab:matrix} reproducible and fair to
the source methods, we fix one transfer setup. Each setting-native
method family is instantiated from its original configuration, preserving
native role structure, team composition, and social mode, and is then run
through the target benchmark's native runner and evaluator. Each family
therefore runs exactly the mechanism it was designed around, including the
within-episode interaction context of its native protocol; cross-episode
memory enters the comparison through the Self-Memory row on the same team.
BBH and HumanEval runs force the social mode
to cooperative because both are cooperative benchmarks, while SOTOPIA,
LLM-Deliberation, and Werewolf retain the source method's native social
mode. Table~\ref{tab:transfer_sources} lists the five source families and
their native protocols.

\begin{table}[tp]
\caption{Source method families for the transfer rows, with native
protocol and benchmark.}
\label{tab:transfer_sources}
\centering
\footnotesize
\setlength{\tabcolsep}{4pt}
\begin{tabular}{p{0.20\linewidth}p{0.46\linewidth}p{0.22\linewidth}}
\toprule
Family & Native protocol & Native benchmark \\
\midrule
ChatDev & Planner$\rightarrow$developer$\rightarrow$tester pipeline (3 agents) & HumanEval \\
AgentNet & 5-agent heterogeneous parallel or routed reasoning & BBH \\
SOTOPIA & Buyer/seller/mediator bargaining (3 agents) & SOTOPIA \\
LLM-Deliberation & 6-party multi-issue proposal voting & LLM-Deliberation \\
Werewolf/DPIns & 5-player hidden-role deception and detection & Werewolf \\
\bottomrule
\end{tabular}
\end{table}

For the cooperative targets (BBH, HumanEval), the source family's role
structure is mapped onto the cooperative runner: a ChatDev pipeline runs as
planner, developer, and tester; an AgentNet or Werewolf team runs as a
five-agent heterogeneous group; and a SOTOPIA trio runs as three
cooperating agents. For SOTOPIA, the source team bargains under the
mixed-motive evaluator. For LLM-Deliberation, the source method's agents
populate the six negotiation parties, with the Werewolf family represented
by the one-adversarial-incentive variant that best matches its structure.
For Werewolf, the source team fills the five player slots under the
adversarial evaluator, except that the fixed five-player count forces
LLM-Deliberation to use its native five-model configuration rather than the
six-party one.

For the Werewolf/DPIns family~\citep{werewolf_dpins}, the original method
trains a discussion policy with reinforcement learning to choose among
discussion tactics, and deduces the other players' hidden roles by belief
modeling rather than by a learned policy; DPIns is the name its released
implementation gives this discussion-policy-instructed agent. To evaluate all
families under the same training-free setting with frozen models, we encode
the discussion tactics it selects as a fixed prompt. This instantiation applies
uniformly across the six evaluation configurations, so the family appears in
every column of the transfer table under the same frozen team.

\paragraph{Team composition.}
Team size and roles follow each benchmark's native protocol: five agents on
BBH, a planner--developer--tester pipeline on HumanEval, a three-role bargaining
setup on SOTOPIA, six parties on LLM-Deliberation, and five players on
Werewolf. The Werewolf team, and the teams used in
the analyses of Appendices~\ref{app:system_diagnostics}
and~\ref{sec:model_heterogeneity}, use the strong lineup, which combines
GPT-4.1, Gemini-2.5-Pro, Gemini-3.5-Flash, GLM-5.2, and Qwen3-14B across three
capability tiers; LLM-Deliberation adds GPT-4o-mini to this lineup for its
sixth party. In the main table, HumanEval uses Gemini-3.5-Flash as planner,
GLM-5.2 as developer, and GPT-4.1 as tester, and SOTOPIA uses GPT-4.1,
Gemini-2.5-Pro, and Gemini-3.5-Flash. BBH uses a weak lineup of five smaller
models of a comparable tier,
where accumulated experience still changes the outcome on a benchmark that a
strong lineup already resolves near its ceiling. Every
main agent is paired with an isolated GPT-4o-mini MemoryAgent. The Werewolf
runs follow the game configuration of the released DPIns
implementation~\citep{werewolf_dpins} without modification; our changes are
confined to the agents that fill the roles.

\paragraph{Evaluation scale.}
Each column of Table~\ref{tab:matrix} is scored over a fixed evaluation set.
BBH uses 100 test questions, five from each of 20 subtasks, following the
protocol of the AgentNet evaluation it is compared against; HumanEval uses all
164 tasks; SOTOPIA uses 80 interactions, eight episodes for each of ten
bargaining scenarios; LLM-Deliberation uses 50 sessions per row; and each
Werewolf audit uses 100 games, above the 30 games per condition used in the
original One Night Ultimate Werewolf study~\citep{werewolf_dpins}. All rows within
a column share the same evaluation protocol.

\paragraph{Sampling precision.}
Each evaluation set is fixed, so the precision of a single entry follows from
its size. In the four columns scored as proportions, the binomial standard
error of one entry is about four points on BBH, under two on HumanEval, and
about five on each Werewolf audit, so a difference of a few points within one
of those columns rests on the paired comparison over the shared items rather
than on the column size alone; Table~\ref{tab:bbh_long} reports such a test.
The LLM-Deliberation column is scored per session and therefore yields an
interval directly: across its 50 sessions the native baseline averages 4.92
(SD 0.90), Self-Memory 4.90 (SD 0.71), and EvolveMAS 5.40 (SD 0.64). The
EvolveMAS advantage of 0.48 over the native baseline and of 0.50 over
Self-Memory each carry a 95\% interval that excludes zero, whether the
sessions are treated as paired or as independent samples, while the 0.02
separating Self-Memory from the native baseline does not.

\paragraph{Determinism and pairing.}
The three BBH rows of Table~\ref{tab:matrix} run under one lineup
configuration and a single fixed seed, and present the tasks in the same
order, so that column is paired item by item. SOTOPIA draws on a fixed set of
ten bargaining scenarios and records no sampling seed; the runs compared there
cover the identical scenario set, so its pairing holds at the scenario level.
The released artifact ships the lineup configuration behind each reported run.

\subsection{Ablation Metric Choice}
\label{sec:ablation_scope}

Each cell of Figure~\ref{fig:ablation} reports one 50-session execution, so
every value is attainable from the sessions behind it. Reading Self-Memory
$K=\varnothing$, Self-Memory$+K_e$, MemoryAgent $K=\varnothing$ and
MemoryAgent$+K_e$, the four conditions order identically on average
agreement (4.90, 5.24, 5.36, 5.40), 5/6 consensus
(78.0, 80.0, 88.0, 90.0), and unanimity on any proposal (24.0, 46.0, 74.0,
86.0). Unanimity on the final proposal (16.0, 34.0, 48.0, 46.0) is the one
exception: MemoryAgent$+K_e$ reaches full agreement on some proposal in
86.0\% of sessions, but marginally less often holds it on the final
proposal than MemoryAgent $K=\varnothing$. The main text
reports unanimity on any proposal because the deliberation task is resolved
only when every party accepts a proposal, and the same ordering holds on
average agreement, the metric carried into Table~\ref{tab:matrix}. This
measure asks whether unanimity occurs at any point in a session, so it is
sensitive to how many proposals a session puts forward; the final-proposal
rate is reported beside it so both readings are visible.

\section{Cross-Setting Transfer Profiles}
\label{sec:transfer_robustness}

Figure~\ref{fig:radar} in the introduction gives a sector-wise comparison of the seven methods
across the six Table~\ref{tab:matrix} configurations. To retain differences
among methods while keeping the heterogeneous metrics legible, the
visualization uses a fixed upper bound of 100 and a configuration-specific
nonzero lower bound $b_j=\tfrac{1}{2}\min_i s_{ij}$. Radial extent is therefore
$(s_{ij}-b_j)/(100-b_j)$; the annotated labels report the percentage-scaled
scores themselves, rounded to one decimal place. Methods should be compared
within each sector rather than by radial-length ratios across sectors. The
resulting profiles make structural transfer visible: cooperative methods are
strongest on the cooperative columns, social methods transfer more naturally
within the mixed-motive region, and adversarial policies concentrate their
strength in the adversarial columns. EvolveMAS is the only method that remains
high across all six columns, consistent with the cross-setting robustness
reported in the main text.

The same profiles show a corresponding structural effect on the cooperative
columns. ChatDev carries a coding-oriented fixed pipeline, SOTOPIA and Deliberation
carry mixed-motive interaction structures, and the Werewolf transfer imposes
the strongest communication constraints; all reach 72.0--74.0 on BBH.
HumanEval is less sensitive to this mismatch, with all methods in a narrow
92.7--97.0 range. Both benchmarks consist largely of independent
tasks, on which a strong cooperative ensemble already performs near ceiling.

\section{MAS-Level Properties}
\label{app:system_diagnostics}

\begin{figure}[tp]
  \centering
  \includegraphics[width=0.84\columnwidth]{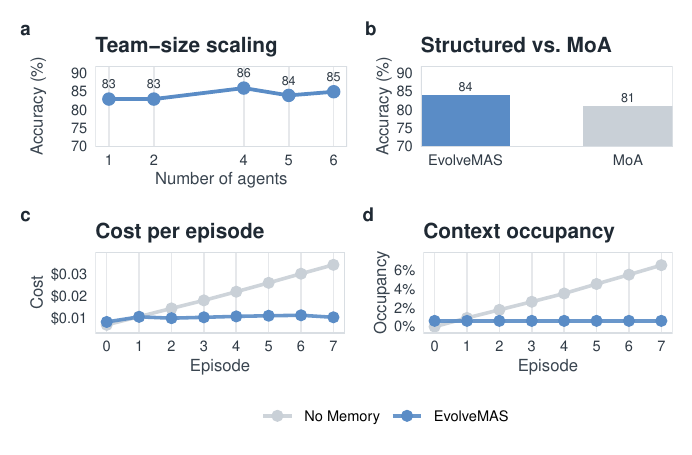}
  \caption{
  Systems stability checks. (a)~Team-size scaling. (b)~Structured cooperation versus static MoA aggregation. (c--d)~SOTOPIA cost and context occupancy over eight episodes.
  }
  \Description{A four-panel systems figure. Panel (a) is a line plot of
  EvolveMAS BBH accuracy at one, two, four, five and six agents,
  remaining in a narrow band. Panel (b) compares EvolveMAS with static
  MoA aggregation. Panels (c) and (d) plot SOTOPIA per-episode cost and
  context occupancy, where EvolveMAS stays flat while No Memory grows.}
  \label{fig:q7}
\end{figure}

This section examines the MAS-level properties of EvolveMAS on traditional
systems metrics, with scalability assessed by non-degradation as the team
grows. \emph{No Memory} denotes an ablation of our own pipeline that runs the
same interaction structure without Structured Memory or Knowledge; in the
multi-episode cost and context checks, its main agents carry the raw
interaction history in context instead. It is therefore reported here and not
as a row of Table~\ref{tab:matrix}. Figure~\ref{fig:q7} addresses three
questions: (a)~whether accuracy is maintained as the team grows;
(b)~whether structured cooperation remains competitive with static
aggregation; and (c--d)~whether amortized cost and
context occupancy remain bounded. Table~\ref{tab:q7_appendix} restates these
checks and adds the short-horizon measurement.

\begin{table}[tp]
\caption{Systems checks. The team-size and aggregation rows
report EvolveMAS only. Cost and occupancy compare EvolveMAS with the
No Memory ablation, which carries the raw interaction history in context.}
\label{tab:q7_appendix}
\centering
\footnotesize
\setlength{\tabcolsep}{3.5pt}
\begin{tabular}{p{0.18\linewidth}p{0.30\linewidth}p{0.42\linewidth}}
\toprule
Question & Check & Result \\
\midrule
Team size\textsuperscript{$\dagger$} &
  BBH acc.\ at 1/2/4/5/6 agents &
  83/83/86/84/85; stable across sizes (Fig.~\ref{fig:q7}a) \\
Aggregation &
  EvolveMAS vs.\ static MoA (5 agents) &
  84.0 vs.\ 81.0 (Fig.~\ref{fig:q7}b) \\
Short horizon &
  GSM8K accuracy &
  94.0 vs.\ No Memory 92.0; maintained \\
Cost &
  SOTOPIA \$/ep at ep 0 / 7 &
  0.0084 / 0.0105 vs.\ No Memory 0.0071 / 0.0341 (Fig.~\ref{fig:q7}c) \\
Context &
  Occupancy at episode 7 &
  0.62\% vs.\ No Memory 6.54\% (Fig.~\ref{fig:q7}d) \\
\bottomrule
\end{tabular}
\end{table}

\noindent\textsuperscript{$\dagger$}\,The team-size sweep holds the
EvolveMAS protocol fixed and varies only the number of main agents. The
weak slot uses Qwen3-14B at 1/2/4 agents and GPT-4.1-mini at 5/6 agents.
At the overlapping 5-agent point, the two weak-slot lineups agree within
$n{=}100$ sampling variation (83--84\%), and point-to-point differences of
1--3\,pp along the sweep lie within the same range.

\paragraph{Team-size scaling.}
Figure~\ref{fig:q7}a plots EvolveMAS BBH accuracy against team size. The
values stay in the 83--86 band across 1, 2, 4, 5, and 6 agents. The
4-agent point of 86.0 lies within $n{=}100$ sampling noise of its
neighbours, so the sweep shows stable accuracy across team sizes, and the
method can be deployed at any of these sizes without re-tuning.

\paragraph{Configuration robustness.}
Figure~\ref{fig:q7}b compares structured cooperation with memory (84.0)
to static Mixture-of-Agents aggregation (81.0) at five agents. Structured
interaction thus compares favorably with pooling independent answers.
Table~\ref{tab:bbh_long} reports the same protocol under the strong and the
weak lineup.

\paragraph{Cost and context.}
Figure~\ref{fig:q7}c--d plot the short SOTOPIA session. EvolveMAS pays a
fixed MemoryAgent overhead and then stays flat; the No Memory ablation
starts cheaper, crosses EvolveMAS within the first
two episodes, and continues to rise as its context accumulates. Context occupancy remains near 0.6\%
for EvolveMAS while No Memory grows with the episode index. The
final-episode values in Table~\ref{tab:q7_appendix} restate the same
pattern: 0.0105 vs.\ 0.0341 in cost, and 0.62\% vs.\ 6.54\% in occupancy.
On the short-horizon GSM8K~\citep{cobbe2021gsm8k} check, EvolveMAS reaches 94.0 against 92.0 for
No Memory, indicating that MemoryAgent preserves performance on independent
single-shot tasks.

\section{Capability-Stratified Sedimentation}
\label{sec:model_heterogeneity}

This section asks whether knowledge accumulation under test-time
self-evolution is uniform across models in a heterogeneous MAS. We examine this
with a per-agent BBH study that inspects the Interaction Knowledge and memory
each model produces under EvolveMAS: rule-level Interaction Strategies and
Semantic Memory entries.

\begin{table}[tp]
\caption{Knowledge sedimentation under EvolveMAS on the heterogeneous BBH team
($n{=}100$ tasks).}
\label{tab:q8}
\centering
\footnotesize
\setlength{\tabcolsep}{4pt}
\begin{tabular}{lcccc}
\toprule
Model (tier) & Sem.\ raw & Sem.\ unique & Rule strategies & Abstraction level \\
\midrule
Gemini-2.5-Pro (strong) & 27 & 11 & 40 & general rules \\
Gemini-3.5-Flash (mid) & 4 & 4 & 40 & subtask features \\
GLM-5.2 (mid) & 11 & 10 & 40 & boundary conditions \\
GPT-4.1 (mid) & 27 & 27 & 40 & procedural advice \\
Qwen3 (weak) & 56 & 11 & 0 & bare answers \\
\bottomrule
\end{tabular}
\end{table}

\begin{figure}[tp]
  \centering
  \includegraphics[width=\columnwidth]{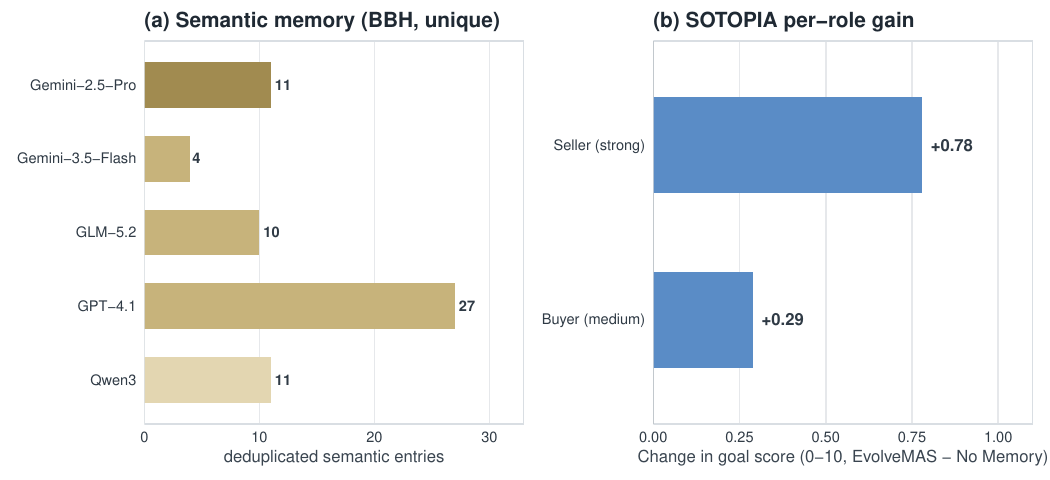}
  \caption{
  Knowledge intake and role-level gains under EvolveMAS. (a)~Unique Semantic Memory entries per model. (b)~Per-role SOTOPIA goal-score change on the 0--10 scale (EvolveMAS $-$ No Memory).
  }
  \Description{A two-panel figure. Panel (a) shows a bar chart of unique
  semantic memory entries for five models, all non-zero, ranging from
  4 to 27. Panel (b) shows per-role SOTOPIA goal-achievement change for a
  strong seller and a medium buyer, both positive, with a larger seller
  gain.}
  \label{fig:q8}
\end{figure}

Table~\ref{tab:q8} shows that the acting model shapes how much transferable
Knowledge an agent distills. Memory intake itself is not gated by capability:
all five models accumulate Semantic Memory entries (column ``Sem.\ unique'').
Distillation into transferable rule-level strategies follows capability: the
four strong/mid models each fill the 40-item strategy store, while the weak
model (Qwen3) distills none. Qwen3's
semantic entries are numerous (56 raw, the highest in the team) but
highly redundant---deduplicated to 11 unique entries, the same count as
Gemini-2.5-Pro---and consist of per-task bare answers rather than
generalizable rules. The gap is therefore in abstraction level, not in
whether memory is produced. Figure~\ref{fig:q8}a shows this intake
directly: every model's semantic store is non-empty, and the weak
model's unique count matches the strongest one's. On the performance
side, a matched per-role SOTOPIA audit under the same evolution protocol
(Figure~\ref{fig:q8}b) shows both roles gaining from EvolveMAS, with the
larger gain on the higher-capacity seller role.

\paragraph{Illustrative case.}
To make the abstraction-level column in Table~\ref{tab:q8} concrete, we
drill into the \texttt{boolean\_expressions} subtask, which all five models attempt under EvolveMAS. Gemini-2.5-Pro (strong) distills the governing
rule directly: \textit{``In boolean logic, the order of operations is
\texttt{not}, then \texttt{and}, then \texttt{or}.''} GLM-5.2 (mid)
captures the same rule plus a boundary case---\textit{``\texttt{not X
and Y} parses as \texttt{not (X and Y)} not \texttt{(not X) and
Y}''}---and a meta-strategy: \textit{``Systematic step-by-step
evaluation respecting precedence is more reliable than left-to-right
reading.''} In Table~\ref{tab:q8} terms, Qwen3's corresponding entries
remain bare answers, with no distilled rules.

\section{Within-Setting Breakdowns}
\label{sec:additional_within_setting}

This section reports the benchmark-specific breakdowns behind the
cross-setting comparison in the main text. The BBH breakdown is
supplementary, while the Werewolf audits define how the WW-V and WW-W
columns of Table~\ref{tab:matrix} are measured.

\subsection{BBH Lineups}

\begin{table}[tp]
\caption{BBH under the weak and the strong lineup. The weak lineup is the
  main-table setup; the strong lineup is the five-model team used in the
  appendix analyses, on which the native AgentNet baseline reaches 84.0\%.
  The transfer families keep their native role structure --- the three-role
  ChatDev pipeline, the SOTOPIA bargaining trio, the five-model
  LLM-Deliberation parallel loop, and the five-player Werewolf/DPIns team ---
  populated with strong-lineup models, and answer the
  same 100 items under the same cooperative loop and highest-confidence
  selection. Answers are scored by option-letter extraction for
  multiple-choice items and exact match on the remaining subtasks. The
  strong-lineup runs form a separate batch from the ablation of
  Table~\ref{tab:ablation_all}.}
\label{tab:bbh_long}
\centering
\footnotesize
\setlength{\tabcolsep}{4pt}
\begin{tabular}{lcc}
\toprule
System & Strong lineup & Weak lineup \\
\midrule
AgentNet (native) & 84.0 & 79.0 \\
Self-Memory & 84.0 & 82.0 \\
EvolveMAS & 82.0 & 83.0 \\
\midrule
ChatDev & 81.0 & 73.0 \\
SOTOPIA & 84.0 & 72.0 \\
LLM-Deliberation & 83.0 & 74.0 \\
Werewolf/DPIns & 87.0 & 72.0 \\
\bottomrule
\end{tabular}
\end{table}

The main-table BBH configuration uses a weaker model lineup,
leaving room to evaluate improvements from accumulated experience.
To examine how these improvements depend on the underlying model
capabilities, an additional evaluation uses a stronger lineup.
This comparison assesses whether the gains persist as baseline
performance approaches the benchmark ceiling.
The stronger-lineup results exhibit a corresponding saturation
pattern, with less room for further improvement.

Table~\ref{tab:bbh_long} contrasts the two lineups ($n{=}100$ each). Under the
weak lineup, EvolveMAS reaches 83.0\%, four points above the 79.0\% native
baseline, with Self-Memory at 82.0\%. Under the strong lineup, Self-Memory
matches the native AgentNet baseline of 84.0\% and EvolveMAS lands two points
below it: a strong ensemble already resolves most BBH instances, so little
experience remains to accumulate across episodes. The lower block places the
cross-setting families on that same strong lineup; every row answers the
identical 100 items, so all comparisons are fully paired. Accuracies span
81--87\%, and exact McNemar tests on the shared items find no significant
difference between any pair of systems except Werewolf/DPIns over ChatDev
($p{=}0.03$, uncorrected for the 21 comparisons). Under the cooperative parallel loop the
LLM-Deliberation family fields the same five-model team as the native
reference, which is why its 83.0 sits beside the 84.0 baseline. We therefore read the strong lineup as
convergent: once a strong ensemble already resolves most BBH items, the
collaboration paradigm a team carries from its source setting no longer
separates it, and little cross-episode structure remains for any paradigm to
exploit.

\subsection{Mechanism Ablation Beyond LLM-Deliberation}
\label{sec:ablation_all}

We ran the same $2\times2$ design on BBH, HumanEval and SOTOPIA
(Table~\ref{tab:ablation_all}). LLM-Deliberation follows Figure~\ref{fig:ablation}; BBH uses the strong
lineup of Table~\ref{tab:bbh_long}; HumanEval uses GPT-4.1 as planner and
developer with Gemini-3.5-Flash as tester; SOTOPIA uses the main-table team
of GPT-4.1, Gemini-2.5-Pro, and Gemini-3.5-Flash. BBH therefore moves from the weak main-table lineup to the strong one, while
the HumanEval ablation team fills every slot with models capable of distilling
rule-level Knowledge (Table~\ref{tab:q8}), because an agent that distills none
leaves the Knowledge axis inert for its share of the team. These lineup
differences are part of why absolute values differ between the ablation and
the main-table audit.
HumanEval reports the pass rate after
up to two repair rounds, and SOTOPIA reports goal achievement averaged over
eight episodes. The LLM-Deliberation row repeats
Figure~\ref{fig:ablation}; the other rows report one designated execution
per condition, separate from the main-table audit and from the BBH runs of
Table~\ref{tab:bbh_long}, so absolute values are not comparable across those
tables. On LLM-Deliberation, BBH and SOTOPIA,
the full configuration (MemoryAgent governance with cross-episode
Knowledge) attains the highest score. BBH is run on the strong lineup, which already
resolves most instances (Table~\ref{tab:bbh_long}) and leaves little
experience to accumulate; the four conditions span two points, inside
the 81--87\% band of Table~\ref{tab:bbh_long}. On
HumanEval, the four conditions differ by at most three of 164 tasks,
which a single run per condition cannot distinguish from noise. On
SOTOPIA, the conditions differ by at most 0.027, of the same size as the
episode-to-episode variation within a single run (standard deviation
0.024--0.047).

\begin{table}[tp]
\caption{
  $2\times2$ factorial on four benchmarks, one designated execution per
  condition. Bold marks the highest value in each row. BBH answers are
  scored as in Table~\ref{tab:bbh_long}.
  }
\label{tab:ablation_all}
\centering
\small
\begin{tabular}{lcccc}
\toprule
& \multicolumn{2}{c}{MemoryAgent} & \multicolumn{2}{c}{Self-Memory} \\
\cmidrule(lr){2-3}\cmidrule(lr){4-5}
Benchmark & $+K_e$ & $K=\varnothing$ & $+K_e$ & $K=\varnothing$ \\
\midrule
LLM-Deliberation & \textbf{5.40} & 5.36 & 5.24 & 4.90 \\
BBH              & \textbf{85.0} & 83.0 & 84.0 & 83.0 \\
HumanEval        & \textbf{97.6} & 96.3 & 95.7 & 97.0 \\
SOTOPIA          & \textbf{0.714} & 0.687 & 0.697 & 0.698 \\
\bottomrule
\end{tabular}
\end{table}

\subsection{Symmetric Werewolf Audit}

\begin{table}[tp]
\caption{
  Werewolf audit ($n{=}100$, heterogeneous team).
  }
\label{tab:werewolf60}
\centering
\footnotesize
\setlength{\tabcolsep}{4pt}
\begin{tabular}{lcc}
\toprule
Method & Village win & Detection \\
\midrule
Self-Memory          & 31.0 & 30.2 \\
EvolveMAS            & \textbf{48.0} & \textbf{44.3} \\
\bottomrule
\end{tabular}
\end{table}

Table~\ref{tab:werewolf60} reports the symmetric audit, which complements
the one-sided evaluations in the main text. Because both sides evolve in this
condition, village win compresses the benefits received by opposing roles
into one zero-sum scalar. We therefore use it as a within-setting summary and
rely on the directional audit in Table~\ref{tab:werewolf_asym} for causal
interpretation. Self-Memory reaches 31.0\%, while EvolveMAS reaches 48.0\%
with a detection accuracy of 44.3\%. The EvolveMAS rate
declines from 0.56 in the first quartile to 0.40 in the final quartile as both
sides accumulate memory, which motivates the directional audit below. Deception success is omitted from this table because under symmetric
evolution it is the complement of village win within the same game, so it
adds no information beyond the first column. Detection accuracy is the rate at
which the village side identifies the wolf, read from the vote distribution
rather than from the win column, so it moves independently of village win.

\subsection{Directional Werewolf Audit}
\label{sec:directional_audit}

Adversarial interactions entail an inherent goal conflict: improving one
side can directly reduce the other side's success. A symmetric deployment
therefore cannot reveal which side benefits from evolution. We therefore run
one-sided conditions that equip the village and wolf sides with EvolveMAS
separately while keeping the opposing side \emph{Unaugmented}: the
benchmark's own Werewolf engine, run without memory and without the DPIns
discussion policy, so neither of its agents is augmented.

\paragraph{One-sided EvolveMAS conditions.}
Table~\ref{tab:werewolf_asym} states all four memory configurations on a
single village-win scale. Village win rises to 0.620 when EvolveMAS is
assigned only to the village and falls to 0.300 when it is assigned only to
the wolf, so the direction of the shift follows the side that receives
EvolveMAS. The two one-sided rows are the same two runs as the WW-V and WW-W
EvolveMAS entries of Table~\ref{tab:matrix}, restated on the village-win
scale rather than each side's own objective: WW-V $=62.0$ is the first one-sided row,
and WW-W $=70.0$ is the corresponding wolf deception success, the complement
of the $0.300$ village win in the second. They are reported here to place both
directions and the two symmetric references on one axis. The
both-sides-evolving reference lies below the Unaugmented reference on this scale,
reflecting the zero-sum compression documented for
Table~\ref{tab:werewolf60}: when village and wolf accumulate memory at the
same time, village win no longer tracks either side's gain, so the one-sided rows carry
the directional claim.

The native DPIns and Self-Memory references in Table~\ref{tab:matrix} remain
close across the two directions. EvolveMAS moves village win toward whichever
side receives it, with the larger shift on the wolf side (from 0.540 to
0.300).

\paragraph{Cross-setting transfer under one-sided roles.}
The transfer rows provide a complementary structural check. Under WW-V,
LLM-Deliberation reaches 64.0, compared with 62.0 for EvolveMAS and 55.0 for
DPIns. Deliberation's multi-party proposal and consensus protocol matches the
village coalition's need to pool distributed clues and converge on a suspect.
Its two-point margin over EvolveMAS corresponds to two games at $n{=}100$, so
we read it as a numerical lead rather than evidence of reliable superiority.
By the same standard, EvolveMAS's own margin over the remaining WW-V rows is
not decisive, and the directional claim rests on WW-W.
The remaining transfer families reach 51.0--55.0.

Under WW-W, EvolveMAS reaches 70.0, compared with 45.0 for AgentNet, 49.0 for
ChatDev, 57.0 for SOTOPIA, 46.0 for LLM-Deliberation, and 44.0 for DPIns.
DPIns's role-agnostic voting instruction is aligned with the village objective,
which explains why, measured on each side's own objective, DPIns reaches 55.0
under WW-V and 44.0 under WW-W. Self-Memory reaches 58.0 under WW-V and 42.0 under WW-W,
and EvolveMAS exceeds it on both sides, most clearly under WW-W (70.0
vs.~42.0).

\begin{table}[tp]
\caption{
  Werewolf memory configurations on a single village-win scale ($n{=}100$ per row). Rows list the village-side policy first.
  }
\label{tab:werewolf_asym}
\centering
\footnotesize
\setlength{\tabcolsep}{8pt}
\begin{tabular}{llc}
\toprule
& Village / wolf configuration & Village win \\
\midrule
\multirow{2}{*}{Symmetric} & Unaugmented / Unaugmented & 0.540 \\
 & EvolveMAS / EvolveMAS & 0.480 \\
\cmidrule(lr){1-3}
\multirow{2}{*}{One-sided} & EvolveMAS / Unaugmented & 0.620 \\
 & Unaugmented / EvolveMAS & 0.300 \\
\bottomrule
\end{tabular}
\end{table}

\section{Long-Horizon Memory Ablation}
\label{app:locomo}

\begin{figure}[tp]
  \centering
\includegraphics[width=\columnwidth]{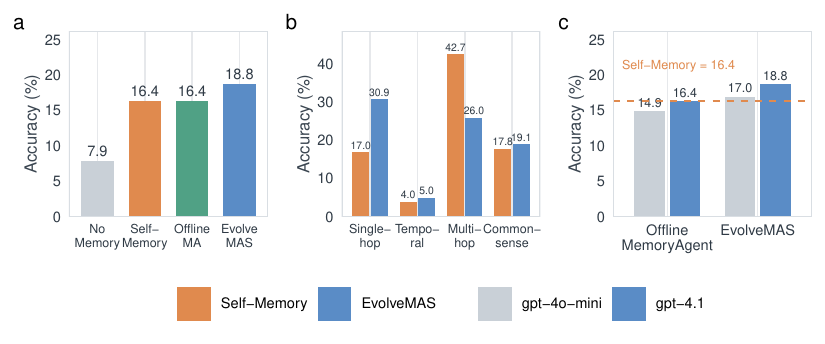}
\caption{
  LoCoMo ablation (gpt-4.1 judge). (a)~Overall accuracy. (b)~Per-category comparison of Self-Memory and EvolveMAS. (c)~Memory model sensitivity, with the dashed line marking Self-Memory.
  }
\Description{A three-panel LoCoMo ablation. Panel (a) shows overall
accuracy bars for No Memory, Self-Memory, Offline MemoryAgent, and EvolveMAS with gpt-4.1 memory, where EvolveMAS
(18.8) exceeds Self-Memory and Offline MemoryAgent (16.4) and No Memory (7.9). Panel (b) shows
per-category grouped bars comparing Self-Memory and EvolveMAS: EvolveMAS leads
on single-hop (30.9 vs 17.0), temporal (5.0 vs 4.0) and common-sense
(19.1 vs 17.8), while Self-Memory leads on multi-hop (42.7 vs 26.0). Panel (c) shows memory model
sensitivity: Offline MemoryAgent and EvolveMAS with gpt-4o-mini vs gpt-4.1, with a dashed Self-Memory
reference at 16.4.}
  \label{fig:locomo}
\end{figure}

The main text reports ownership through the benchmark-native Self-Memory
and EvolveMAS pairs. This section adds a LoCoMo~\citep{maharana2024locomo} ablation that examines memory
management from the memory perspective, separating memory ownership from
update timing. LoCoMo evaluates question answering over long multi-session
conversations and allows the starting memory to be held fixed, which isolates
the effect of when the MemoryAgent updates. Figure~\ref{fig:locomo} reports
the ablation.

Each condition answers the same 1,540 questions, the full set of ten
conversations at 154 questions each. A gpt-4.1 judge scores every answer as
correct or incorrect, counting an answer correct when it carries the reference
information whatever its wording, and a deterministic filter applied ahead of
the judge marks an answer incorrect when it declines to commit --- ``not
mentioned'', ``not specified'' --- against a reference that states a concrete
fact. Absolute accuracies are therefore not comparable with published LoCoMo
results, which apply the benchmark's released protocol to answers produced
outside a multi-agent loop. Every arm here shares this judge and this filter,
so the comparison among them is internally consistent, and the grading program
ships with the released code.

Holding the starting memory fixed and varying only whether the MemoryAgent
updates during the session, online updates (EvolveMAS) beat offline batch
updates (Offline MemoryAgent) overall (18.8 vs.~16.4 with gpt-4.1 memory).
The gain is concentrated on single-hop questions (30.9 vs.~24.8), where fresh
per-question facts are most reusable, and the two are tied on multi-hop
questions (26.0 each), where an offline batch has more opportunity to
cross-link distant sessions before answering. These per-category values for
Offline MemoryAgent are reported here only; Figure~\ref{fig:locomo}b contrasts
Self-Memory with EvolveMAS. Relative to
No Memory, which answers from the most recent session alone (7.9), both
Offline MemoryAgent and EvolveMAS show substantial memory gains
($+$8.5--10.9\,pp). EvolveMAS also exceeds Self-Memory overall (18.8
vs.~16.4) and leads on fact recall (single-hop 30.9 vs.~17.0, common-sense 19.1
vs.~17.8), while Self-Memory is strongest on multi-hop reasoning (42.7
vs.~26.0), where its in-context self-notes link distant sessions.
Self-Memory and Offline MemoryAgent land on the same overall value, so what
separates the arms on this benchmark is online updating: EvolveMAS improves on
both static arms by the same 2.4 points.

\section{Qualitative Cases}
\label{sec:qualitative}

To make the self-evolution mechanism concrete, we show representative
memory, skill, and profile artifacts produced under EvolveMAS. The content is taken from the Structured Memory and Interaction Knowledge
stores written during evaluation and lightly trimmed for length. These cases show
that the same memory substrate---episodic, semantic, and working
regions---is reused across all six benchmark columns. What changes is the stored
content. In cooperative tasks, semantic entries emphasise role-specific
observations and coordination cues, while skill artifacts emphasise reusable
procedures. In mixed-motive tasks, the substrate retains partner signals and
negotiation anchors, while episode-end artifacts expose compact partner models
or tactics. In adversarial tasks, it retains deception-relevant cues
that remain valid across games. This
setting-adaptive content helps explain why the main-table result is more than a
collection of separate mechanisms.

\begin{membox}{BBH (cooperative reasoning) -- evolved collaboration skill, EvolveMAS, GPT-4.1}
Skill: strategy_collaborate_cooperative   type: collaboration  (stored under EvolveMAS)
Agent: agent_gpt41   Model: GPT-4.1   Setting: cooperative (BBH, n=100)
When to use: Always -- this is the agent's current evolved role.

Rule: In cooperative mode, always propose my initial deductions but
explicitly invite my collaborator to check for logical consistency and
point out any contradictions before finalizing the answer.
\end{membox}

The strategy is short, actionable, and reusable across BBH subtasks. It
stores a verification rule rather than a task-specific answer. That is the
kind of coordination regularity the paradigm is meant to distill.

\begin{membox}{HumanEval (cooperative coding) -- verification-focused working memory, EvolveMAS}
Working (tester, episode HumanEval/9):
  focus = verify rolling_max against edge cases
  next  = confirm candidate([])==[] and monotone prefix invariants
\end{membox}

In the cooperative coding line, the reusable signal is a role-specific test
focus rather than a cached final answer. The stored state tells the tester
what to verify next. Reuse is therefore anchored to boundary-case checking
and task decomposition.

\begin{membox}{SOTOPIA (mixed-motive) -- negotiation anchors, EvolveMAS}
Example episode: Antique Chair bargaining

Episodic:
  - listed an Antique Chair at $15
  - buyer offered $9 -> countered at $10
  - buyer accepted $10 -> pickup arranged
Semantic:
  - item_value = The Antique Chair is valued at $15 but can be negotiated down to $10 (high)
  - interaction_regularity = expressing flexibility in pricing can facilitate negotiation (medium)
Working:
  focus: Sell the Antique Chair for at least $10
  next:  Maintain a friendly negotiation tone to build rapport with the buyer
\end{membox}

Here the reusable content is a compact negotiation anchor set: an acceptable
price floor and a simple interaction regularity. The interaction regularity extends beyond this item, and the price anchor
guides the current negotiation. They help explain how
EvolveMAS reaches the native SOTOPIA goal-achievement level.

\begin{membox}{LLM-Deliberation (mixed-motive) -- stance flexibility, EvolveMAS}
Agent: SportCo   Condition: EvolveMAS   example session reaching 5/6 agreement

Episodic (deal trajectory, one session):
  - proposed A1,B1,C4,D1,E5  -> blocked by Local Labour Union
  - Local Labour Union proposed A2,B3,C3,D2,E3
  - Mayor, Environmental League, Dept. of Tourism supported A2,B3,C3,D2,E3
  - SportCo proposed A2,B3,C3,D2,E3 (switched, did not re-assert blocked deal)
  - final agreement reached on A2,B3,C3,D2,E3   (5/6 parties agree)
\end{membox}

This trace shows how EvolveMAS updates a stance during deliberation. Once
SportCo's opening deal is blocked, it moves to the coalition-preferred deal,
and the session reaches agreement among five of the six parties. This
behavioral pattern helps explain the high consensus rates of EvolveMAS.

\begin{membox}{Werewolf (adversarial) -- transferable cues, EvolveMAS, n=100, VWR=0.48}
Agent: player2   Condition: EvolveMAS
Model: GPT-4.1 (heterogeneous team)   overall VWR = 0.48   early-session = 0.56, late = 0.40 (zero-sum)

Profile and semantic state:
  player1_tendency = has been voted for and revealed not to be a Werewolf in one observed game
  player1_reliability = low
  role_swap_rule = Robber can swap roles with another player, but call order matters
  call_order_cue = stories that violate the night-action order are suspect

\end{membox}

These entries retain reusable cues: the night-action order and the role-swap
rule of the game as reproduced here~\citep{werewolf_dpins}, which expose claims
that violate them. The same accumulation loop that distills a coordination
rule in BBH, a negotiation anchor in SOTOPIA, and a stance shift in
Deliberation thus yields durable rule-level cues in Werewolf. Under
symmetric evolution, village win starts at 0.56 early in the session and
settles at 0.40 as both sides accumulate memory under zero-sum dynamics, and
the one-sided audits in Table~\ref{tab:werewolf_asym} attribute the shift to
the side that receives EvolveMAS. The unified substrate therefore relies on one
memory-governance procedure that adapts what it stores to what remains true
across resets.